\documentclass[aps,prc,preprint,superscriptaddress,showpacs]{revtex4}
\usepackage[dvips]{graphicx}
\usepackage{amsmath,amssymb}

\def\delsla{\!\not\!\partial}
\begin{document}
\preprint{SAGA-HE-234-07}
\title{Effects of a multi-quark interaction on color superconducting 
phase transition in an extended NJL model}

\author{Kouji Kashiwa}
%\email[]{kashiwa2scp@mbox.nc.kyushu-u.ac.jp}
\affiliation{Department of Physics, Graduate School of Sciences, Kyushu University,
             Fukuoka 812-8581, Japan}
             
\author{Masayuki Matsuzaki}
\email[]{matsuza@fukuoka-edu.ac.jp}
\affiliation{Department of Physics, Fukuoka University of Education, 
             Munakata, Fukuoka 811-4192, Japan}

\author{Hiroaki Kouno}
%\email[]{kounoh@cc.saga-u.ac.jp}
\affiliation{Department of Physics, Saga University,
             Saga 840-8502, Japan}

\author{Masanobu Yahiro}
%\email[]{yahiro2scp@mbox.nc.kyushu-u.ac.jp}
\affiliation{Department of Physics, Graduate School of Sciences, Kyushu University,
             Fukuoka 812-8581, Japan}

\date{\today}

\begin{abstract}
 We study the interplay of the chiral and the color superconducting 
phase transition in an extended Nambu--Jona-Lasinio model with a multi-quark 
interaction that produces the nonlinear chiral-diquark coupling.
We observe that this nonlinear coupling adds up coherently 
with the $\omega^2$ interaction 
to either produce the chiral-color superconductivity coexistence phase or 
cancel each other depending on its sign. 
We discuss that a large coexistence region in the phase diagram is consistent 
with the quark-diquark picture for the nucleon whereas its smallness is 
the prerequisite for the applicability of the Ginzburg-Landau approach. 
\end{abstract}

\maketitle

 The findings of recent ultra-relativistic heavy ion collision experiments 
have stimulated a paradigm shift, that is, quark gluon plasma (QGP) is not a 
weakly interacting near ideal gas but a strongly interacting near perfect 
fluid, called sQGP, at least slightly above the transition 
temperature~\cite{Lee,GM,Shu}.
Quantum chromodynamics (QCD) exhibits a variety of forms of matter also at 
high density; chiral symmetry restoration, deconfinement, and color 
superconductivity (CSC)~\cite{RW}. 
As for CSC, various sub-phases at intermediate density are discussed 
recently in addition to standard two flavor superconductivity and 
color flavor locking. 

 First principle lattice QCD simulations describe high temperature phenomena 
but their applicability to finite density is limited due to the well known 
sign problem and/or zero eigen values of the fermion matrix. 
Therefore effective models, such as the Nambu--Jona-Lasinio (NJL) 
model~\cite{NJ1,NJ2,Kle,HK1} 
or the random matrix model~\cite{Ste,Hal}, must be employed. 
Recently the Polyakov-NJL model that handles not only chiral restoration but 
also deconfinement is actively studied~\cite{MO,Fuk,Rat,Ghos}. 

 Before CSC came into consideration, standard effective models predicted 
that chiral restoration at high density is a first order transition. 
(See also Ref.~\cite{Sak} and references cited therein that go beyond the mean 
field approximation.) 
However, the vector interaction, which is not forbidden from symmetry 
consideration, rather necessary from a view point of nuclear physics but often 
ignored, may change the situation~\cite{Bu,Kit}. The competition between the chiral 
$\langle\bar qq\rangle$ 
and the diquark $\langle qq\rangle$ condensates on the temperature-chemical 
potential ($T$-$\mu$) plane was first considered by Berges and Rajagopal~\cite{BR}. 
In their calculation, the two types of condensates are mutually exclusive. 
But, in principle, they can coexist, {\it i.e.}, quarks dressing chiral condensate 
can pair up. 
Pairing between such 
constituent quarks would lead to the quark-diquark picture 
for the nucleon~\cite{LBT}. On the other hand, 
recently Hatsuda et al. obtained interesting results including a new end point 
induced by the U(1) anomaly in the three flavor case, 
using a model-independent 
Ginzburg-Landau (GL) approach to the $T$-$\mu$ phase diagram~\cite{Hat}. 
Here it should be 
noted that the GL approach is applicable when all 
the order parameters considered are small, 
since the free energy is expanded with respect to them. 
In the present case, both $\sigma$ and $\Delta$ should be small. 
This situation is realized in the vicinity of each phase transition, 
if the coexistence region is small or vanishes. 
It is thus an important information how large the region is, 
and effective models are useful to answer the question. 

 We adopt a framework that can handle these aspects of the QCD phase diagram 
on the same footing --- an extended NJL model with multi-quark interactions. 
(For introduction of multi-quark interactions, see also Osipov et 
al.~\cite{Osi1,Osi2} and Huguet et al.~\cite{Hug1,Hug2}.) In a previous 
paper~\cite{Kas1}, we found that the $\sigma^2\omega^2$ 
and the $\sigma^4$ interactions sharpen the chiral transition weakened 
by the $\omega^2$ interaction and also 
that the $\sigma^4$ interaction shifts the 
critical end point to a higher $T$, lower $\mu$ point. In the present paper, 
we discuss the interplay of the chiral and the diquark condensates brought about 
by the $\sigma^2\Delta^2$ interaction and its coherence/competition with the 
$\omega^2$ interaction. Here, $\sigma$, $\omega$, and $\Delta$ denote the 
scalar, vector, and diquark auxiliary mean fields, respectively, defined later. 
Note that our study is limited to the two flavor case at present. 

The Lagrangian density of the extended NJL model adopted in the present work 
is given by
\begin{eqnarray}
  {\cal L}&=&{\bar q}( i \delsla - m_0 )q
            + \Bigr[{g}_{2,0} \Bigl( 
               (\bar q q)^2 + (\bar q i \gamma_5  \vec{\tau} q)^2 \Bigl)
              +{g}_{4,0}\Bigl( 
               (\bar q q)^2 + ( \bar q i\gamma_5  \vec{\tau} q)^2 \Bigr)^2 
                                                          \nonumber \\
          &&  -{g_{0,2}}( \bar q \gamma^\mu q)^2
              -{g_{2,2}} \Bigl( (\bar q q)^2  
              +( \bar q i\gamma_5  \vec{\tau} q)^2 \Bigr)
	       (\bar q \gamma^\mu q)^2 \nonumber \\ 
          &&  +{d_{0,2}} \Bigl(
               (i{\bar q}^c \varepsilon \epsilon^b \gamma_5 q)
                  (i{\bar q} \varepsilon \epsilon^b \gamma_5 q^c)
	      +(i{\bar q}^c \varepsilon \epsilon^b \gamma_5 q)
                  (i{\bar q} \varepsilon \epsilon^b \gamma_5 q^c)
               \Bigr) \nonumber\\
         &&   +{d_{2,2}} (\bar q q)^2\Bigl(
               (i{\bar q}^c \varepsilon \epsilon^b \gamma_5 q)
                  (i{\bar q} \varepsilon \epsilon^b \gamma_5 q^c)
	      +(i{\bar q}^c \varepsilon \epsilon^b \gamma_5 q)
                  (i{\bar q} \varepsilon \epsilon^b \gamma_5 q^c)
               \Bigr)
              + \cdots \Bigr] ,  
\end{eqnarray}
where $q^c=C{\bar q}^T$ and ${\bar q}^c=q^TC$ are the charge-conjugation spinors,
$C=i \gamma^2 \gamma^0$ is the charge-conjugation matrix, 
$q$ is the two flavor quark field,
${\vec \tau}=(\tau^1,\tau^2,\tau^3)$ are the Pauli matrices,
$m_0={\rm diag}(m_u,m_d)$ is the current quark mass matrix,
$\varepsilon$ and $\epsilon^b$ are the totally antisymmetric tensors 
in the flavor and color spaces, 
and $g_{i,j}$ and $d_{m,n}$ ($i,j,m,n=0,1,2,\cdots$) are 
the coupling constants of quark-quark interactions.
We consider only the four- and eight-quark interactions ignoring 
higher-order interactions denoted by ellipsis in the Lagrangian density. 
Among the physically important eight-quark interactions, 
$g_{4,0}$ and $g_{2,2}$ terms will be ignored in the following, since 
their roles have been clarified in our previous paper~\cite{Kas1} as mentioned 
above and we would like to concentrate on the chiral-diquark coexistence. 
Under the standard mean field approximation (MFA), the Lagrangian density reads
\begin{eqnarray}
{\cal L}_{\rm MFA} &=& {\bar q}( i \delsla - (m_0+\Sigma_{\rm s}) 
                                           + \Sigma_{\rm v} \gamma^0)q 
             - \frac{1}{2} \Sigma_{\rm d}^{*b}
                         (i{\bar q}^c \varepsilon \epsilon^b \gamma_5 q)
             - \frac{1}{2} \Sigma_{\rm d}^b 
                         (i{\bar q} \varepsilon \epsilon^b \gamma_5 q^c)
             - U ,
\end{eqnarray}
where
\begin{eqnarray}
\Sigma_{\rm s}  &=& -2( {g}_{2,0} \sigma
                   + d_{2,2} \sigma |\Delta|^2 ) 
                   ,\nonumber\\
\Sigma_{\rm v}  &=& - 2{g}_{0,2} \omega 
                   ,\nonumber\\
\Sigma^b_{\rm d} &=& -2( d_{0,2} \Delta^b 
                             + d_{2,2} \sigma^2 \Delta^b) 
                   ,\nonumber\\
\Sigma^{*b}_{\rm d} &=& -2( d_{0,2} \Delta^{*b} 
                                + d_{2,2} \sigma^2 \Delta^{*b})
                   ,\nonumber\\
U&=& g_{2,0}\sigma^2 
        -g_{0,2}\omega^2 
        + d_{0,2}
          \Delta^{*b} \Delta^b
        + 3 d_{2,2} \sigma^2\Delta^{*b} 
          \Delta^b ,
\end{eqnarray}
and the auxiliary fields introduced are 
$\sigma = \langle {\bar q}q \rangle$,
$\omega = \langle {\bar q}\gamma^0 q \rangle$,
$\Delta^b = \langle i{\bar q}^c \varepsilon \epsilon^b \gamma_5 q \rangle$, and 
$\Delta^{*b} = \langle i{\bar q}^c \varepsilon \epsilon^b \gamma_5 q \rangle$.

 The thermodynamical potential $\Omega$ of the system 
with finite temperature $T$ and chemical potential $\mu$
is then obtained as
\begin{eqnarray}
\Omega &=& -2 N_f V 
                 \Bigl[\int \frac{d^3{\bf p}}{(2 \pi)^3} E_{\bf p} 
                       +\frac{1}{\beta} \Bigl\{ \ln(1+e^{-\beta E_{\bf p}^+})
                 + \ln(1 +e^{-\beta E_{\bf p}^-}) \Bigr\} \nonumber \\
	 && +{\rm sgn}(E_{\bf p}^-)E^-_\Delta + E^+_\Delta
            +\frac{2}{\beta} \Bigl\{ 
             \ln(1+e^{-{\rm sgn}(E_{\bf p}^-)\beta E^-_\Delta})
             +\ln(1+e^{-\beta E^+_\Delta})
             \Bigr\}
                 \Bigr] \nonumber \\
         && + V U , 
\end{eqnarray}
where $\beta = 1/T$, $\tilde{\mu} = \mu + \Sigma_{\rm v}$, $M = m_0+\Sigma_{\rm s}$, 
$E_{\bf p} = \sqrt{{\bf p}^2+M^2}$, 
$E_{\bf p}^\pm = E_{\bf p} \pm \tilde{\mu}$, 
$E^\pm_\Delta = \sqrt{E_{\bf p}^{{\pm}2}+|\Sigma_{\rm d}|^2}$, 
and 
${\rm sgn}(E_{\bf p}^-)$ is the sign function.
The corresponding scalar, vector, and scalar diquark densities,
$\rho_{\rm s},~\rho_{\rm v}$ and $\rho_{\rm d}$ are given by
\begin{eqnarray}
\rho_{\rm s} &=& -2 N_f M \int \frac{d^3p}{(2\pi)^3}
           \frac{1}{E_{\bf p}}
           \Bigl\{
             1-n(E^-_{\bf p}) -n(E^+_{\bf p})
             + \frac{E^-_{\bf p}}{E^-_\Delta} \tanh \frac{\beta E^-_\Delta}{2}
             + \frac{E^+_{\bf p}}{E^+_\Delta} \tanh \frac{\beta E^+_\Delta}{2}
           \Bigr\} ,
\end{eqnarray}
\begin{eqnarray}
\rho_{\rm v} &=& 2 N_f \int \frac{d^3 p}{(2 \pi)^3}
           \Bigl\{
                   n(E^-_{\bf p}) - n(E^+_{\bf p})
                  -\frac{E^-_{\bf p}}{E^-_\Delta}
                     \tanh \frac{\beta E_\Delta^-}{2}
                  +\frac{E^+_{\bf p}}{E^+_\Delta}
                     \tanh \frac{\beta E_\Delta^+}{2}
           \Bigr\} ,
\end{eqnarray}
\begin{eqnarray}
\rho_{\rm d} &=& -2 N_f \Sigma_d \int \frac{d^3 p}{(2 \pi)^3}
           \Bigl\{
                   \frac{1}{E_\Delta^-}\tanh \frac{\beta E_\Delta^-}{2}
                  +\frac{1}{E_\Delta^+}\tanh \frac{\beta E_\Delta^+}{2}
           \Bigr\} ,
\end{eqnarray}
where 
\begin{eqnarray}
n_q=\frac{1}{1+\exp{\{\beta(E_p-\tilde{\mu})\}}},~~~~~
n_{\bar q}=\frac{1}{1+\exp{\{\beta(E_p+\tilde{\mu})\}}} .
\end{eqnarray}
The gap equation can be derived by minimizing 
the thermodynamical potential with respect to 
$\sigma$, $\omega$, and $\Delta^*$, 
their physical solutions then satisfy the stationary condition
\begin{eqnarray}
\left(
\begin{array}{c}
{\partial \over{\partial \sigma}}\left({\Omega\over{V }}\right) \\
{\partial  \over{\partial \omega}}\left({\Omega\over{V }}\right)\\
{\partial  \over{\partial \Delta^*}}\left({\Omega\over{V}}\right)
\end{array}
\right)
&=&{\cal G}
\left(
\begin{array}{c}
\sigma -\rho_{\rm s} \\
\rho_{\rm v}-\omega \\
\Delta - \rho_{\rm d}
\end{array}
\right)
=\left(
\begin{array}{c}
0 \\
0 \\
0
\end{array}
\right), 
\end{eqnarray}
where the effective couplings are 
\begin{eqnarray}
{\cal G}&\equiv& 
\left(
\begin{array}{ccc}
G_{{\rm s}\sigma} & G_{{\rm v}\sigma} & G_{{\rm d} \sigma} \\
G_{{\rm s}\omega} & G_{{\rm v}\omega} & G_{{\rm d} \omega} \\
G_{{\rm s}\Delta} & G_{{\rm v}\Delta} & G_{{\rm d} \Delta} \\
\end{array}
\right) \nonumber \\
&\equiv&
\left(
\begin{array}{ccc}
-{\partial \Sigma_{\rm s}\over{\partial \sigma}} & 
-{\partial \Sigma_{\rm v}\over{\partial \sigma}} & 
-{\partial \Sigma_d^*\over{\partial \sigma}} \\
-{\partial \Sigma_{\rm s}\over{\partial \omega}} & 
-{\partial \Sigma_{\rm v}\over{\partial \omega}} & 
-{\partial \Sigma_d^*\over{\partial \omega}} \\
-{\partial \Sigma_{\rm s}\over{\partial \Delta^* }} & 
-{\partial \Sigma_{\rm v}\over{\partial \Delta^* }} & 
-{\partial \Sigma_d^*\over{\partial \Delta^* }} \\
\end{array}
\right) \nonumber \\
&=&
\left(
\begin{array}{ccc}
2(g_{2,0}+d_{2,2}|\Delta|^2) & 
0 & 
4d_{2,2}\sigma \Delta^* \\
0 & 
2g_{0,2} & 
0 \\
4d_{2,2}\sigma \Delta & 
0 & 
2(d_{0,2}+d_{2,2}\sigma^2) \\
\end{array}
\right) .
\end{eqnarray}
When $\det({\cal G}) \neq 0$, ${\cal G}$ has its inverse, 
and then the stationary condition leads to
$\sigma=\rho_{\rm s}$, $\omega=\rho_{\rm v}$, and $\Delta = \rho_{\rm d}$. 

It has been shown that the effect of the $\omega^2$ coupling 
on the phase diagram 
is suppressed by the non-linear terms,
$g_{4,0}\sigma^4$ and $g_{2,2}\sigma^2\omega^2$, in our previous paper~\cite{Kas1}. 
Therefore, the vector coupling $g_{0,2}$ is fixed to the small value, $0.2g_{2,0}$.
The adopted parameters for numerical calculations are summarized in Table~I. 
We examine both signs for $d_{2,2}$ since this is not determined within the model 
and they would lead to different physical pictures. 

\begin{table}[h]
\begin{center}
\begin{tabular}{lllllll}
\hline
model 
& $g_{2,0}$ & $g_{0,2}$ 
& $d_{0,2}$ & $|d_{2,2}| \sigma_0^2$ 
\\
\hline
NJL + $\Delta^2$ 
%     & 10.9960215~~~
& 5.498~~~~
& 0~~~
& $G_{\Delta}$~~~~
& 0~~~
\\
NJL + $\omega^2$ + $\Delta^2$ 
%     & 10.9960215~~~
& 5.498
& $ G_\omega$
& $G_{\Delta}$
& 0~~~
\\
NJL + $\Delta^2$ + $\sigma^2 \Delta^2$ 
%     & 10.9960215~~~
& 5.498
& 0~~~
& $G_{\Delta}$
& $0.2 G_{\Delta}$
\\
NJL + $\omega^2$+$\Delta^2$ + $\sigma^2 \Delta^2$ 
%     & 10.9960215~~~
& 5.498
& $ G_\omega $
& $G_{\Delta}$
& $0.2 G_{\Delta}$
\\
\hline
\end{tabular}
\caption[]{Summary of the parameter sets. The coupling constants are shown in 
${\rm  GeV}^{-2}$. For all cases we adopt $m_0=0.0055~{\rm GeV}$,
$\Lambda=0.6315~{\rm GeV}$, and 
$\sigma_0=-0.03023~{\rm GeV}$.
Here, $G_\omega$ and $G_\Delta$ are  
$0.2 g_{2,0}$ and $0.6 g_{2,0}$, respectively. (See Table~1 in Ref.~\cite{Kas1} for 
comparison.)
}
\end{center}
\end{table}

\begin{figure}[htbp]%[H]
\begin{center}
 \includegraphics[width=7.5cm]{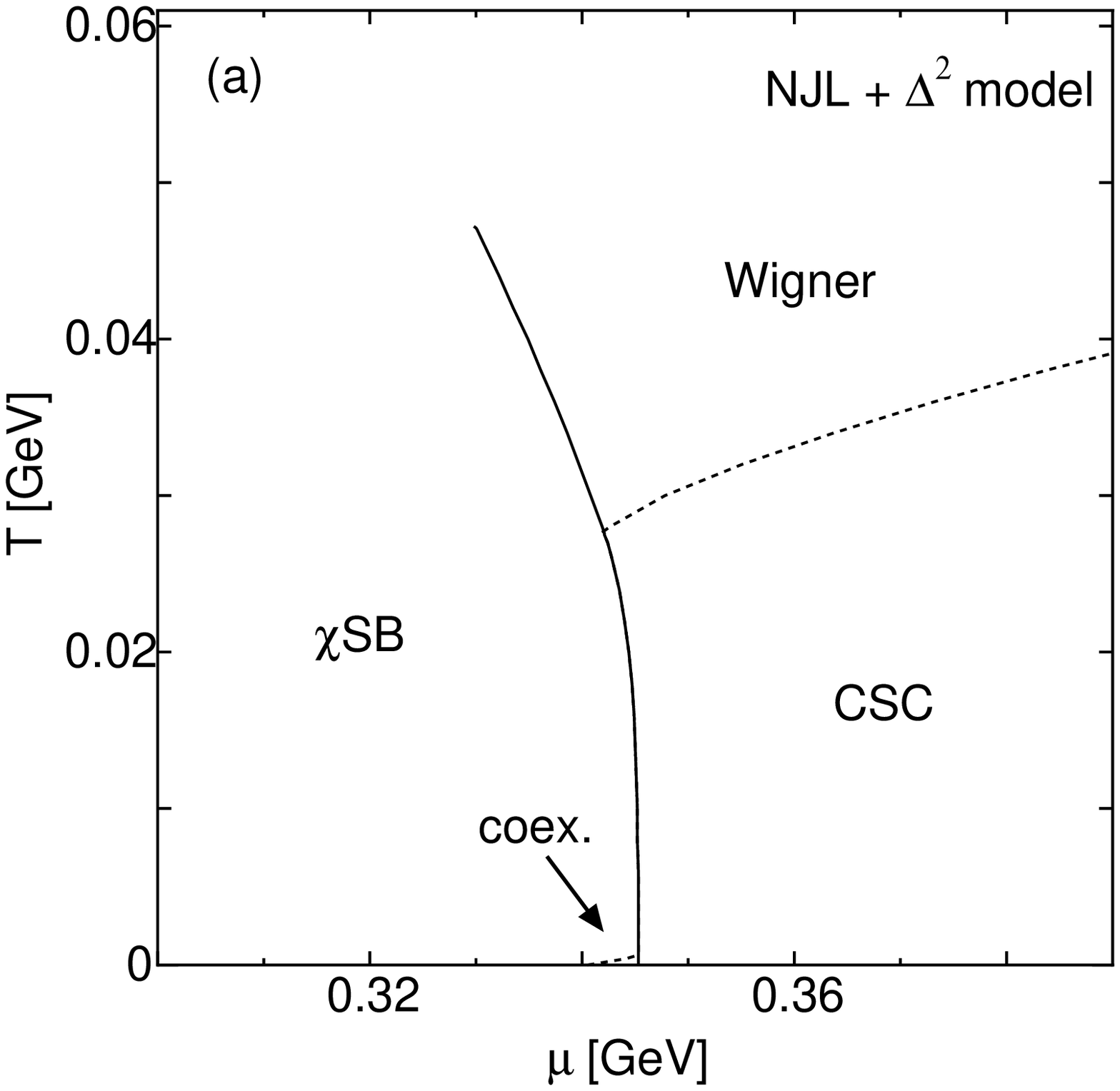} 
 \includegraphics[width=7.5cm]{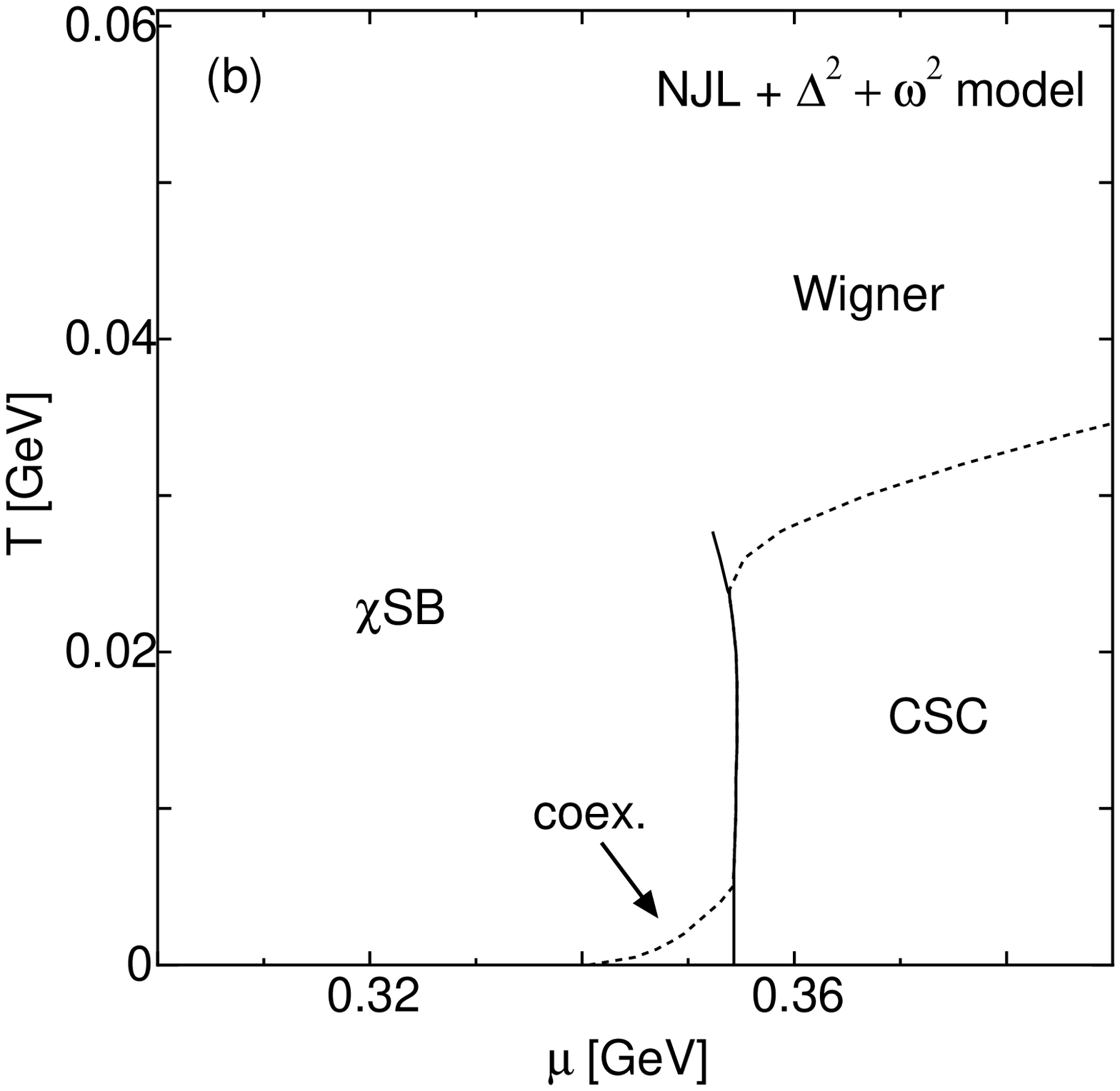}
\end{center}
\caption{Phase diagrams given by (a) the standard linear NJL model and (b) the 
extended NJL model that includes the $\omega^2$ interaction.
}
\label{fig1}
\end{figure}

\begin{figure}[htbp]%[H]
\begin{center}
 \includegraphics[width=7.5cm]{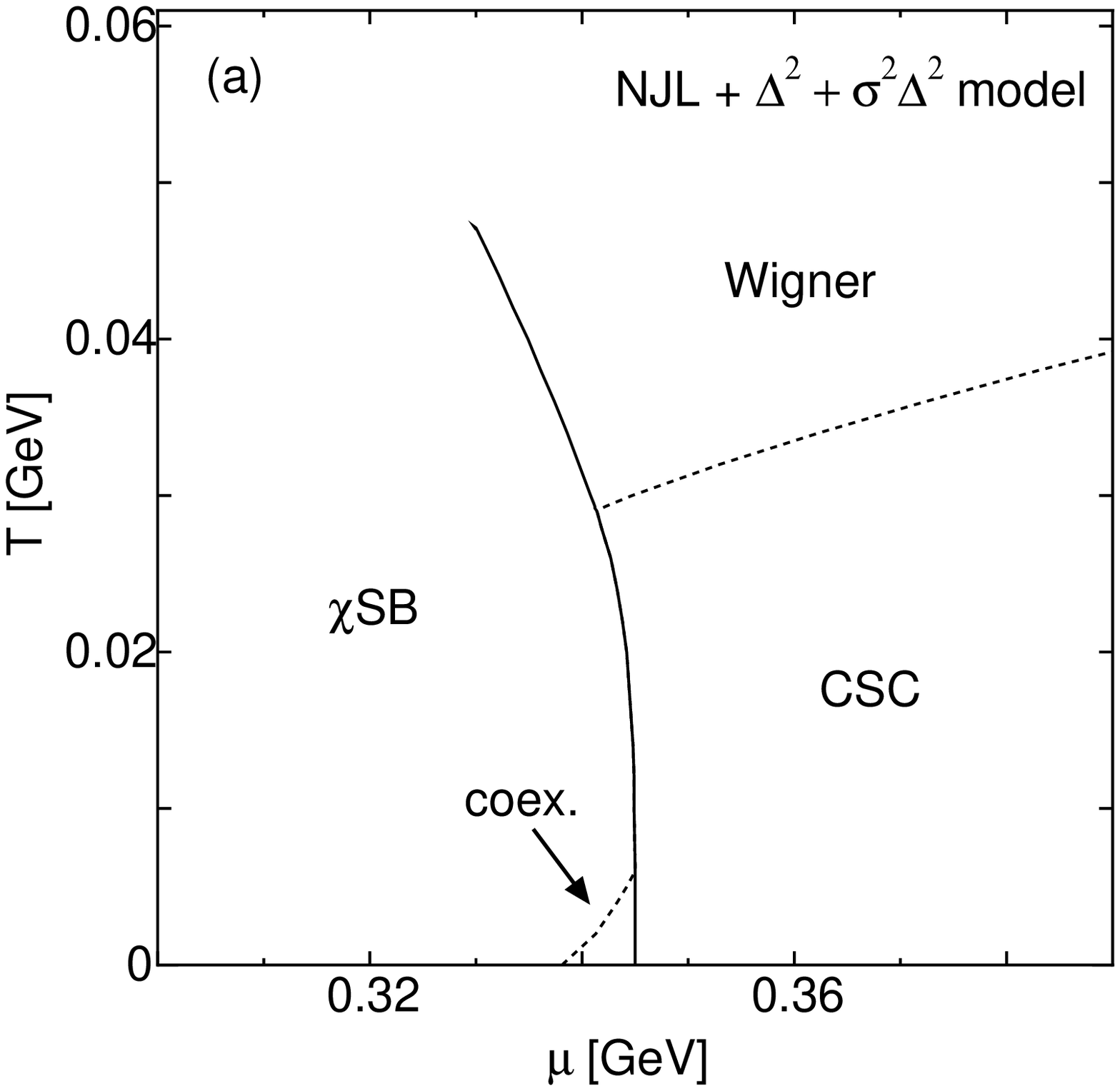} 
 \includegraphics[width=7.5cm]{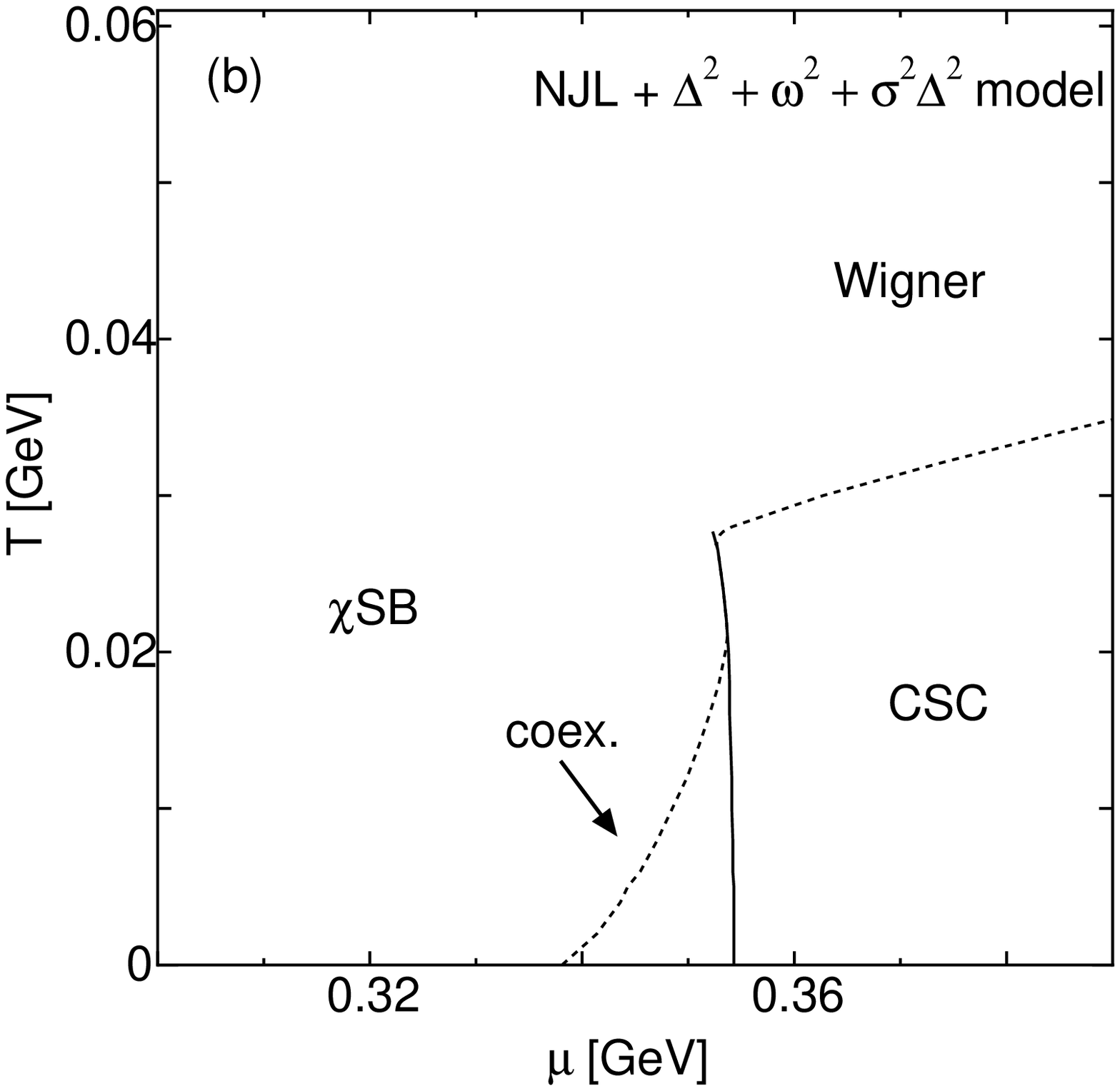}
\end{center}
\caption{Phase diagrams given by (a) the extended NJL model that includes the 
$\sigma^2\Delta^2$ interaction and (b) that includes the $\omega^2$ and the 
$\sigma^2\Delta^2$ interactions. In these calculations $d_{2,2}$ is positive. 
}
\label{fig2}
\end{figure}

\begin{figure}[htbp]%[H]
\begin{center}
 \includegraphics[width=7.5cm]{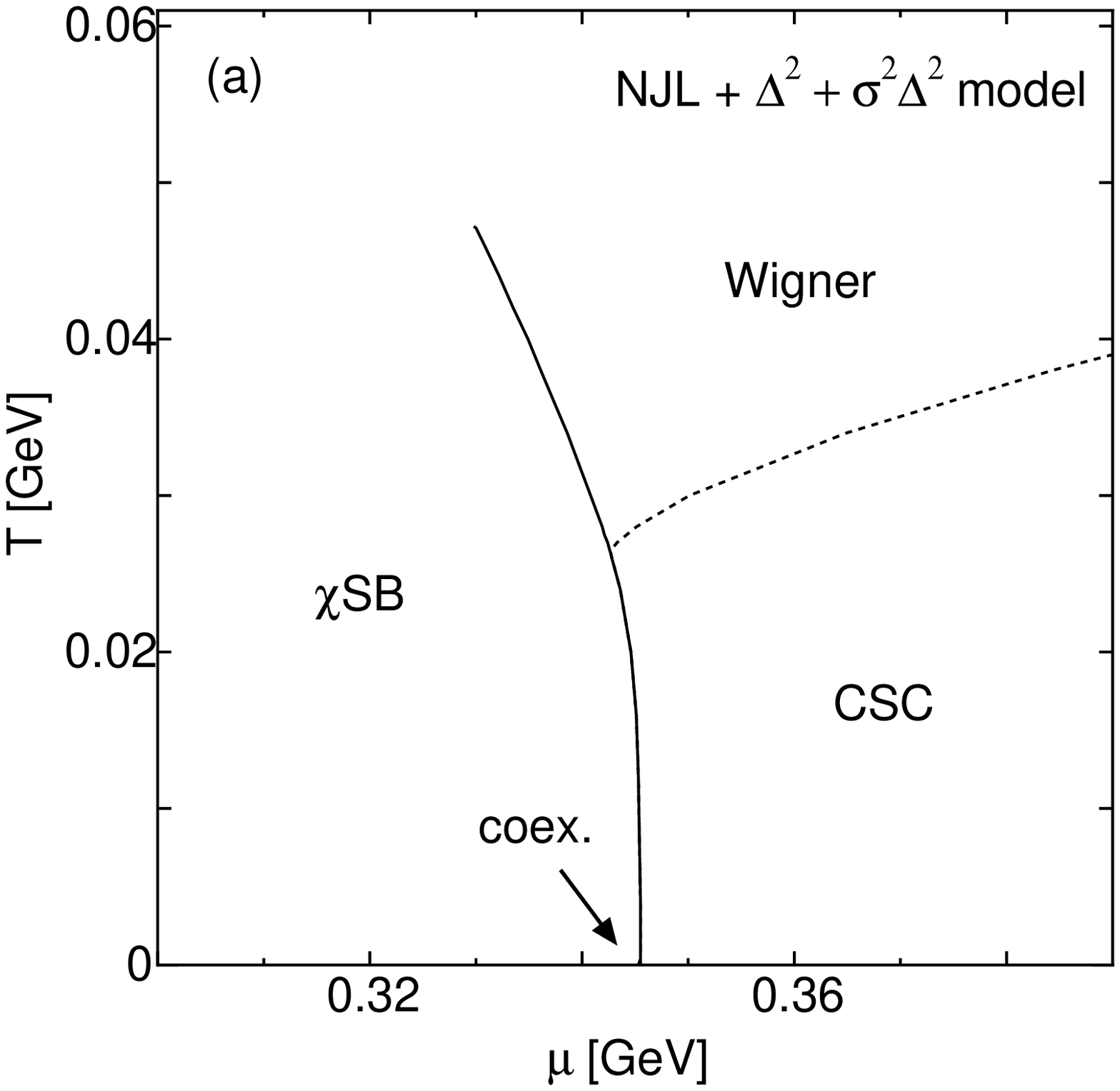} 
 \includegraphics[width=7.5cm]{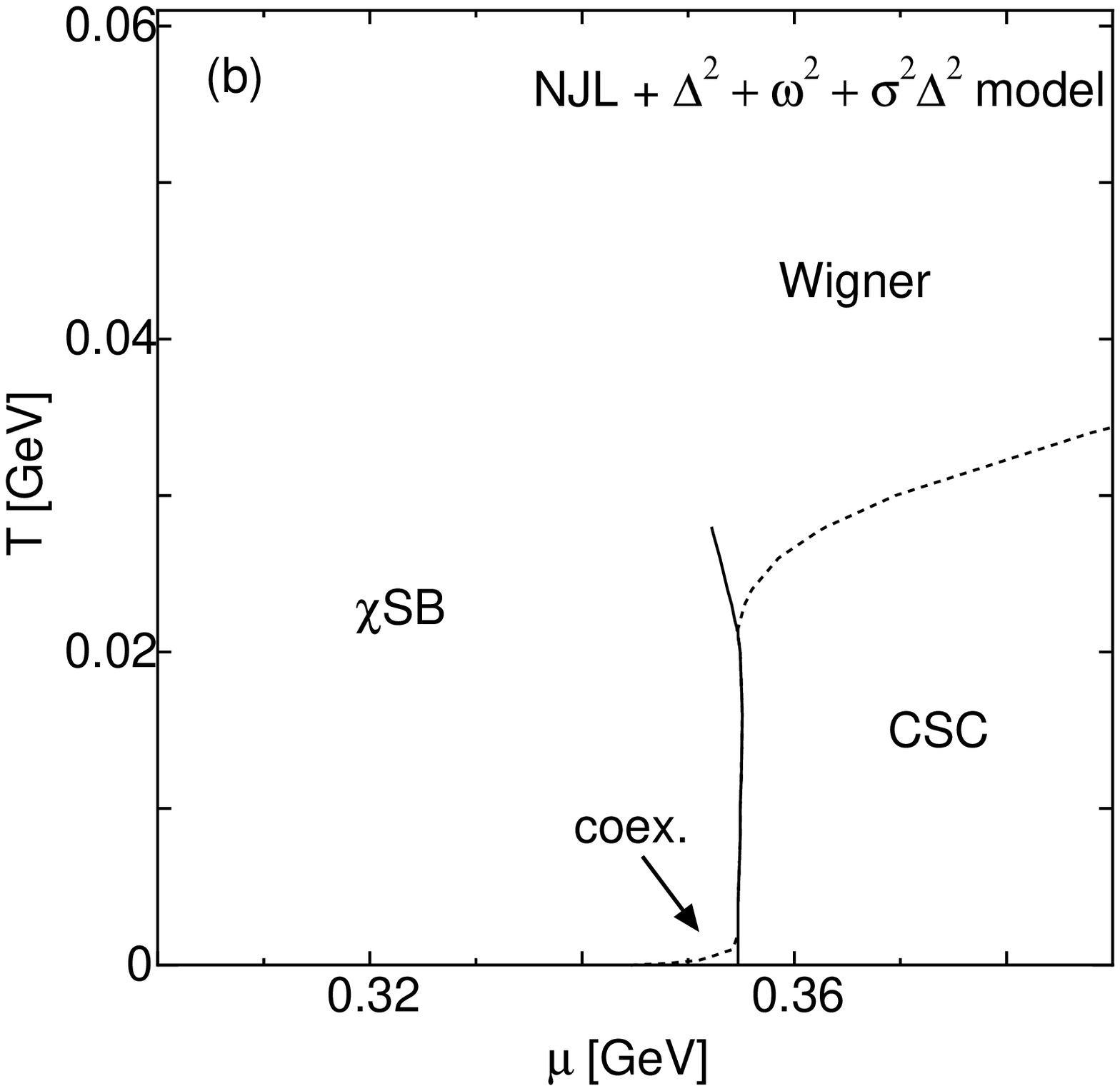}
\end{center}
\caption{The same as Fig.~\ref{fig2} but $d_{2,2}$ is negative.
}
\label{fig3}
\end{figure}

 In the following, we discuss the phase diagrams obtained 
 by adopting the models 
with the parameters summarized in Table~I putting emphasis on the 
chiral-diquark coexistence at low-$T$. 
Figure~\ref{fig1}(a) graphs the phase diagram of the standard NJL model with 
the diquark condensate. The coexistence region is very small in this case. 
Blaschke et al. first pointed out the existence of this coexistence region 
adopting another parameter set without the $\omega^2$ interaction~\cite{BVY}. 
Figure~\ref{fig1}(b) shows the effect of 
the $\omega^2$ interaction, that is, it weakens 
both transitions and shifts the chiral restoration to the higher density side 
and consequently produces a coexistence phase at low-$T$. This confirms 
the result presented by Kitazawa et al.~\cite{Kit}. 
Comparison of Figs.~\ref{fig1}(a) and \ref{fig2}(a) demonstrates the effect of 
the $\sigma^2\Delta^2$ coupling for the case of positive $d_{2,2}$. 
This nonlinear interaction shifts the CSC 
transition to lower density and consequently produces a coexistence phase. 
Figure~\ref{fig2}(b) includes both the $\omega^2$ and the $\sigma^2\Delta^2$ 
interactions. They coherently add up in this case. 
This result can be understood from the expressions of the effective couplings 
that lead to
\begin{equation}
\Sigma_{\rm s}=-G_{{\rm s}\sigma}\sigma=-2\Bigl({g}_{2,0}+d_{2,2}|\Delta|^2\Bigr)\sigma
\end{equation}
and
\begin{equation}
\Sigma_{\rm d}=-G_{{\rm d}\Delta}\Delta=-2\Bigl(d_{0,2}
                             +d_{2,2}\sigma^2\Bigr)\Delta ,
\end{equation}
in which 
the former indicates that positive $d_{2,2}$ enhances $|\Sigma_{\rm s}|$ when 
$\Delta\neq0$ exists and the latter indicates that positive $d_{2,2}$ enhances 
$|\Sigma_{\rm d}|$ when $\sigma\neq0$ exists. 
This means that the $\sigma^2\Delta^2$ interaction acts 
only when the $\omega^2$ 
interaction makes $\sigma$ and $\Delta$ coexisting. 
Figures~\ref{fig3}(a) and 
\ref{fig3}(b) graph the result of the negative $d_{2,2}$. In this case the 
$\omega^2$ interaction and the $\sigma^2\Delta^2$ one are destructive to each 
other and consequently the coexistence region becomes very small. 

%%%%%%%%%%%%%%%% Fig 1 %%%%%%%%%%%%%%%%%%%%% ->
%%%%%%%%%%%%%%%%%%%%%%%%%%%%%%%%%%%%%%%%% 

%%%%%%%%%%%%%%%% Fig 2 %%%%%%%%%%%%%%%%%%%%% ->
%%%%%%%%%%%%%%%%%%%%%%%%%%%%%%%%%%%%%%%%% 

%%%%%%%%%%%%%%%% Fig 3 %%%%%%%%%%%%%%%%%%%%% ->
%%%%%%%%%%%%%%%%%%%%%%%%%%%%%%%%%%%%%%%%% 

 As mentioned above, the sign and the magnitude of the eight-quark interaction 
is not determined within the present model. 
The positive $d_{2,2}$ results in a large coexistence region. The sign is 
supported by the result of the quark-diquark model for the nucleon~\cite{LBT} 
that the diquark interaction is sizably stronger in the 
normal baryon-number density region than in the high density one. 
However, further analysis is needed to determine the strength of 
the coupling more precisely in the normal density region.
Oppositely, if the negative sign is favored by some 
reason, the coexistence region 
shrinks and the two phase transitions occur at 
almost the same $T$ and $\mu$. In the vicinity, 
the order parameters, $\sigma$ and $\Delta$, are small. 
Such a region is an ideal 
playground for the GL approach, since the free energy is 
expanded with respect to them. 
In other words, when the sign is positive, 
the GL model is not useful to determine the phase diagram except for 
the high-$T$ region where both the order parameters are small. 

 We have studied the interplay of the chiral and the color superconducting 
phase transition in an extended Nambu--Jona-Lasinio model with a multi-quark 
interaction that produces the nonlinear $\sigma^2\Delta^2$ coupling. 
We have found that the size of the chiral-diquark coexistence 
region is sensitive to the sign of the coupling. 
The positive sign is supported by the quark-diquark model for 
the nucleon, but further analysis is 
needed to determine the density dependence of 
the diquark interaction more precisely. 
Meanwhile, the negative sign is the prerequisite for the 
applicability of the Ginzburg-Landau approach that has already been 
applied to determine the phase diagram. 
Thus, the determination of the sign is an important subject related to 
the phase diagram.

\end{document}